\documentclass[a4paper]{article}
\usepackage{INTERSPEECH2021,amsmath,graphicx}

\usepackage{times}
\usepackage{epsfig}
\usepackage{amssymb}
\usepackage{microtype}
\usepackage{booktabs}
\usepackage[caption=false,font=footnotesize]{subfig}
\usepackage{pifont}
\usepackage{relsize}
\usepackage{multirow}
\usepackage{gensymb}
\usepackage{tabularx}
\usepackage{comment}
\usepackage{xcolor}
\usepackage{amsmath}
\usepackage{cite}
\usepackage{etoolbox}

\title{Multi-Channel Transformer Transducer for Speech Recognition}
\name{Feng-Ju Chang, Martin Radfar, Athanasios Mouchtaris, Maurizio Omologo}
\address{Alexa Machine Learning, Amazon, USA} 
\email{\tt\{fengjc, radfarmr, mouchta, omologo\}@amazon.com}

\begin{document}
\maketitle
\begin{abstract}
Multi-channel inputs offer several advantages over single-channel, to improve the robustness of on-device speech recognition systems. Recent work on multi-channel transformer, has proposed a way to incorporate such inputs into end-to-end ASR for improved accuracy. However, this approach is characterized by a high computational complexity, which prevents it from being deployed in on-device systems. In this paper, we present a novel speech recognition model, \emph{Multi-Channel Transformer Transducer (MCTT)}, which features end-to-end multi-channel training, low computation cost, and low latency so that it is suitable for streaming decoding in on-device speech recognition. 
In a far-field in-house dataset, our MCTT outperforms stagewise multi-channel models with transformer-transducer up to $6.01\%$ relative WER improvement (WERR). 
In addition, MCTT outperforms the multi-channel transformer up to $11.62\%$ WERR, and is $15.8$ times faster in terms of inference speed. 
We further show that we can improve the computational cost of MCTT by constraining the future and previous context in attention computations.  
\end{abstract}


\section{Introduction}
\label{sec:intro}

Voice assisted devices nowadays are usually equipped with multiple microphones for far-field speech recognition in noisy environments~\cite{haeb2020far,vincent2018audio}. By combining the spectral and spatial information of target and interference signals captured from different microphones, the beamforming approaches ~\cite{omologo2001speech,wolfel2009distant,kumatani2012microphone,kinoshita2016summary,virtanen2012techniques,menne2016rwth} have been demonstrated to benefit automatic speech recognition (ASR) systems substantially for improved recognition accuracy \cite{barker2015third,kinoshita2016summary,menne2016rwth}. The beamformer thus has become the standard module, typically introduced before the ASR front-end and acoustic model.

The delay-and-sum and super-directive beamformers~\cite{doclo2007superdirective,himawan2010clustered} are among the most popular beamforming methods for ASR, the latter one characterized by both its higher directivity and its lack of robustness to imperfect microphone arrays~\cite{chen2021robustness}.
With the great success of deep neural networks, neural beamformers have gained significant interest and are becoming the state-of-the-art technologies in end-to-end all-neural ASR systems~\cite{heymann2016neural,erdogan2016improved,ochiai2017multichannel,chang2019mimo,chang2020end,kumatani2019multi,li2016neural,xiao2016deep,meng2017deep,liu2014using}.
The neural beamforming methods are generally categorized into fixed beamforming (FBF)~\cite{kumatani2019multi,liu2014using} and adaptive beamforming (ABF) methods~\cite{heymann2016neural,erdogan2016improved,ochiai2017multichannel,chang2019mimo,chang2020end,li2016neural,meng2017deep} depending on whether the beamforming weights are fixed or varied based on the input signals during inference time.

While neural beamforming approaches are attractive for their model capacity and direct access to the downstream ASR loss for optimizing the beamforming weights, their performance is still hindered by stagewise training. For example, the neural mask estimators in ABF methods~\cite{heymann2016neural,erdogan2016improved} usually must be pre-trained on synthetic data where the target speech and noise labels are well defined. The mismatch of these statistics between synthetic data and real-world data, however, can lead to noise leaking into the target speech statistics~\cite{drude2019unsupervised}, and deteriorate its finetuning with the cascaded acoustic models. 

Bypassing the need for stage-wise optimization and leveraging the core ability of transformer networks~\cite{vaswani2017attention}, i.e. attention on multiple modalities, a single integrated multi-channel transformer network was proposed~\cite{chang2021end} with both channel-wise and cross-channel attention layers for joint beamforming and acoustic modeling. Despite its effectiveness, this model is hard to apply to the streaming case such as on-device speech recognition~\cite{he2019streaming}, which demands low latency and low computation. First, it relies on an attention mechanism (encoder-decoder attention) over full encoder outputs to learn alignments between input and output sequences~\cite{bahdanau2014neural}. Second, the input audio is encoded in a bidirectional way, thus requiring a full utterance as input. Furthermore, the attention computation increases quadratically with the length of input sequences. Finally, the model size of the multi-channel transformer increases w.r.t. the number of microphones and the number of time frames~\cite{chang2021end} due to the use of affine transformations to aggregate multi-channel embeddings in cross-channel attention layers. For these reasons, it is unsuitable for on-device ASR systems with small memory.

There exist many streamable ways for alignment learning such as connectionist temporal classification (CTC)~\cite{graves2006connectionist}, transducer~\cite{graves2012sequence}, monotonic chunkwise attention (MoChA)~\cite{chiu2017monotonic}, and triggered attention~\cite{moritz2019triggered}, all of which can be integrated with transformer~\cite{vaswani2017attention,dong2018speech,lu2020exploring,wang2020transformer,gulati2020conformer}. In this work, we focus on transducer due to its outstanding performance over traditional hybrid models for streaming speech recognition~\cite{he2019streaming,sainath2020streaming}. Several research efforts have combined transformer with transducer for single-channel speech recognition~\cite{tian2019self,yeh2019transformer,zhang2020transformer,huang2020conv}, but to the best of our knowledge, it is the first time that transducer is integrated with multi-channel transformer.

In addition to achieving streamable alignment learning, we further make the encoders streamable via limiting future context (right-context) and previous context (left-context) in both channel-wise and cross-channel attention computations for multi-channel audio encoding, and constraining previous context in self attention for output sequence embedding as well. For cross-channel attention computations, we also propose to use two simple combiners, the average and concatenation of multiple channels to create keys and values. In this way, our model size does not increase as the number of microphone and input sequence length increase.

\begin{figure*}[t]
\centering
\includegraphics[width=1.0\textwidth]{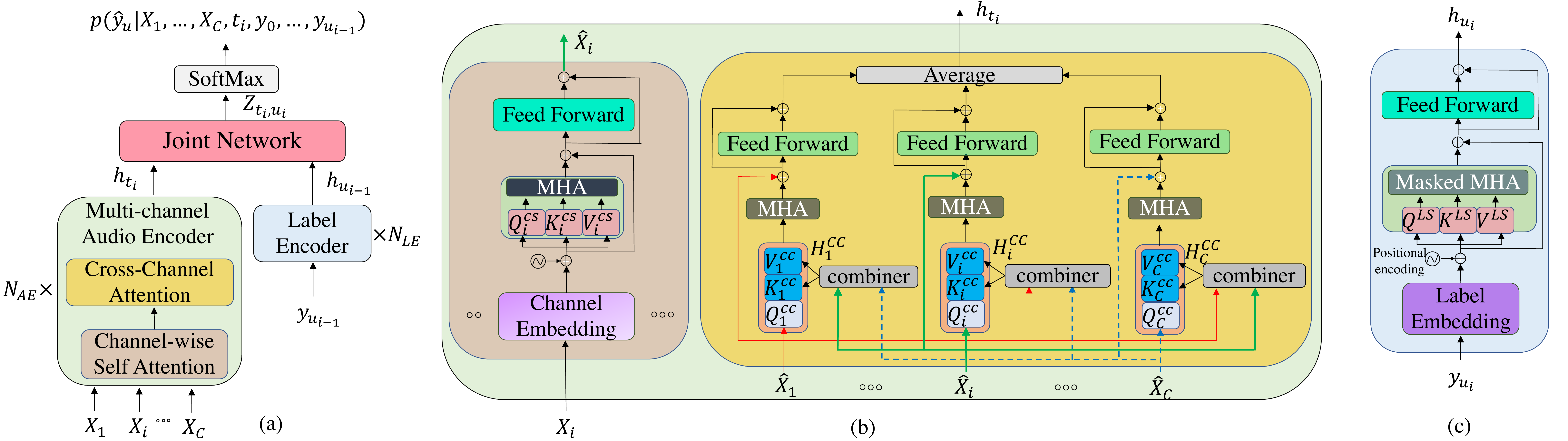}
\caption{An overview of the multi-channel transformer transducer (MCTT). (a) The high-level block diagram of MCTT (b) The multi-channel audio encoder architecture, which contains $N_{AE}$ channel-wise attention layers (left) and cross-channel attention layers (right). MHA represents multi-head attention, and $C$ is the number of channels. (c) The label encoder architecture, which consists of $N_{LE}$ self-attention layers with token labels as inputs. Note that the layer norm is applied in both MHA and feed-forward layers, but omitted here.}
\vspace{-4mm}
\label{fig:model_diagram}
\end{figure*}

In a far-field in-house dataset, we show that the proposed multi-channel transformer transducer outperforms single channel and stagewise neural beamformers cascaded with transformer transducers by $7.14\%$ and $6.01\%$ WERR respectively. Moreover, our model performs better than multi-channel transformer~\cite{chang2021end} up to  $11.62\%$ WERR and is $15.8$ times faster in terms of inference speed (TP50). Finally, we improve the computational cost of both multi-channel audio encoder and label encoder for streaming case, by limiting both the left and right context in attention computations. Moreover, the performance gap between the causal attention and full attention versions of our model can be bridged by attending to a limited number of future frames.

\section{Multi-Channel Transformer Transducer}
\label{sec:method}

\subsection{Transducer}

We denote $C$-channel of audio sequences as $\mathcal{X}=(X_1,...,X_i,...,X_C)$ where each channel is of $T$ frames, $X_i=(\mathbf{x}_{i,1},\mathbf{x}_{i,2},...,\mathbf{x}_{i,T})$. We also denote a transcription label sequence of length $U$ as $\mathbf{y}=(y_1,y_2,...,y_U)$, where $y_u \in \mathcal{Z}$, and $\mathcal{Z}$ is a predefined set of token labels. As depicted in Fig.~\ref{fig:model_diagram} (a), the transducer model encodes acoustic sequences first with a multi-channel audio encoder network (Fig.~\ref{fig:model_diagram} (b)) to produce encoder output states as $h=(h_1,...,h_T)$. For each encoder state $h_t$, the model predicts either a label or a blank symbol $\langle b \rangle$ with a joint network. If the model predicts a blank symbol, which indicates the lack of token label for that time step, then the model proceeds to the next encoder state. Different from CTC~\cite{graves2006connectionist}, the transducer model exploits not only the encoder output at time $t$ but also the previous non-blank label history as inputs to predict the next output. The previously predicted labels are encoded with a label encoder as shown in Fig.~\ref{fig:model_diagram} (c).

The transducer model defines a conditional distribution,
\begin{align}
    P(\hat{\mathbf{y}}|\mathcal{X}) = \prod_{i=1}^{T+U} P(\hat{y}_i|\mathcal{X},t_i,y_0,...,y_{u_{i-1}})
    \label{eq:cond_dist}
\end{align}
where $\hat{\mathbf{y}}=(\hat{y}_1,...,\hat{y}_{T+U}) \subset \{\mathcal{Z} \cup \langle b \rangle\}^{T+U}$ correspond to any possible alignment path with $T$ blank symbols and $U$ labels such that after removing all blank symbols in $\hat{\mathbf{y}}$ yields $\mathbf{y}$, and $y_0$ is the start of sentence symbol.

We can marginalize $P(\hat{\mathbf{y}}|\mathcal{X})$ over all possible alignments $\mathcal{A}(\mathcal{X},\mathbf{y})$ to obtain the probability of the target label sequence $\mathbf{y}$ given the input multi-channel sequences $\mathcal{X}$,
\begin{align}
    P(\mathbf{y}|\mathcal{X}) = \sum_{\hat{\mathbf{y}} \in \mathcal{A}(\mathcal{X},\mathbf{y})} P(\hat{\mathbf{y}}|\mathcal{X})
    \label{eq:marg_dist}
\end{align}
This alignment probability summation can be computed efficiently with forward-backward algorithm~\cite{graves2012sequence}.

\begin{figure*}[t]
\centering
\includegraphics[width=0.95\textwidth]{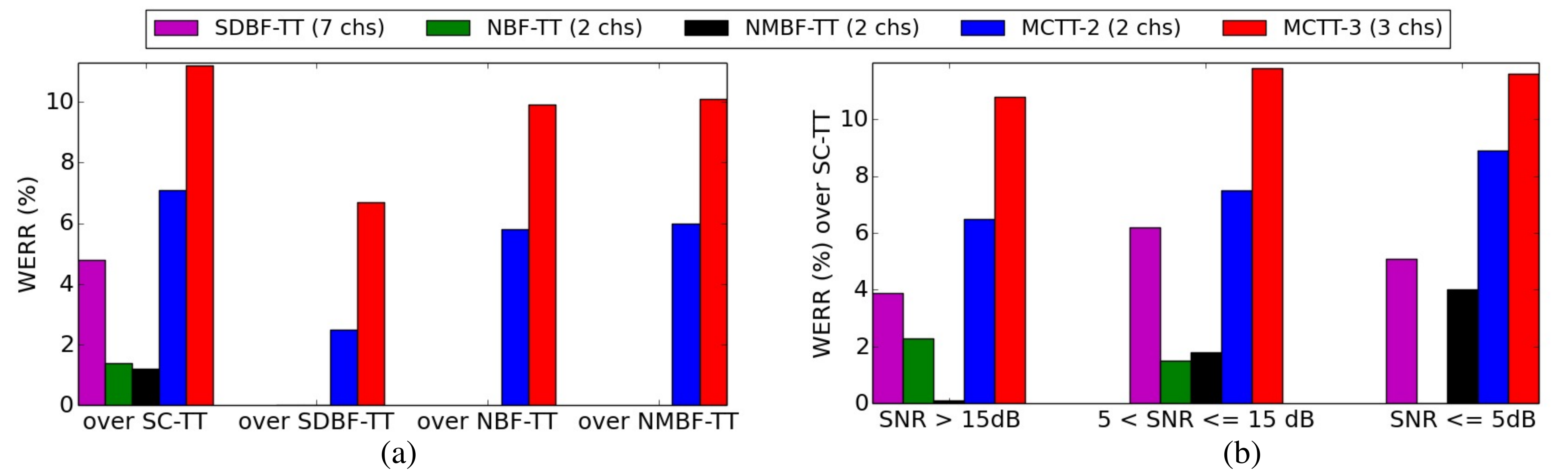}
\vspace{-10pt}
\caption{The relative word error rate reduction, WERRs (\%), by comparing the multi-channel transformer transducer (MCTT) to the beamfomers cascaded with transformer transducers. (a) WERRs over different methods (b) WERRs over SC-TT w.r.t. different SNR levels. A higher number indicates a better WER. Negative WERRs are not reported.}
\label{fig:werrs}
\end{figure*}

\subsection{Multi-Channel Audio Encoder}
\label{sec:MCEnc}
Previous work on the transducer framework~\cite{graves2012sequence,tian2019self,yeh2019transformer,zhang2020transformer,huang2020conv} relied only on single-channel input.
To address multi-channel inputs, we propose to build our audio encoder based on multi-channel transformer network~\cite{chang2021end}, as shown in Fig.~\ref{fig:model_diagram} (b), containing two main blocks, channel-wise self-attention layers and cross-channel attention layers.

\textbf{Channel-wise Self-Attention Layer (CSA)}: We start by projecting the source channel features (log-STFT magnitude and phase features are used in this work) to the dense embedding space for more discriminative representations. Then the embedded features plus the positional encoding~\cite{vaswani2017attention} are fed into a set of learnable weight parameters to create Query ($Q_i^{CS}$), Key ($K_i^{CS}$), Value ($V_i^{CS}$). Similar to~\cite{chang2021end}, the transformed features, $Q_i^{CS}$ and $K_i^{CS}$, are used to compute the correlation across time steps within a channel via multi-head attention (MHA)~\cite{vaswani2017attention}. The resulting attention matrix is then used to reweight the features of $V_i^{CS}$ in each time step followed by a feed-forward network to produce the self-attention outputs.

\textbf{Cross-Channel Attention Layer (CCA)}: Given the self-attended outputs per channel, the cross-channel attention layers aim to learn the contextual relationship across channels both within and across time steps. Inspired by~\cite{chang2021end}, when we use the $i$-th channel to create $Q_i^{CC}$, the other channels are leveraged by a combiner to create $K_i^{CC}$ and $V_i^{CC}$. Different from~\cite{chang2021end} which takes the sum of channel encodings after applying affine transformations (\emph{Affine}), we investigate two simple combiners: (1) \emph{Avg}: take the average of the other channels along both time and embedding axes, $H_i^{CC}=1/C\sum_{j \neq i} \hat{X}_j$, which can be seen as the symmetric weight case of the \emph{Affine} combiner in \cite{chang2021end} (2) \emph{Concat}: concatenate the other channels along the time axis, $H_i^{CC} = [\hat{X}_1;...;\hat{X}_j;...;\hat{X}_C]_{j \neq i}$. Here, $\hat{X}_j \in \mathbb{R}^{T \times d}$ and $d$ is the embedding size. With this adaptation, the model parameters do not increase w.r.t. the number of microphones ($C$) and time frames ($T$) as in~\cite{chang2021end}. Finally, the cross-channel attention outputs are fused by a simple average.

\subsection{Label Encoder and Joint Network}
We leverage the transformer network to build the label encoder, as illustrated in Fig.~\ref{fig:model_diagram} (c). An embedding layer converts previously predicted non-blank labels into vector representations. Then several linear layers project the embedding vectors in order to create $Q^{LS}$, $K^{LS}$, and $V^{LS}$ followed by masked MHA computations. The attention scores from the future frames are always masked out to ensure causality. Note that label encoder outputs do not attend to multi-channel audio encoder outputs, in contrast to the architecture in~\cite{chang2021end}. As discussed in Sec.~\ref{sec:intro}, doing so poses a challenge for streaming applications. Instead, we use a joint network, which is a fully-connected feed-forward neural network with a single hidden layer and $tanh$ as the activation function. We concatenate outputs of multi-channel audio encoder and label encoder as inputs to the joint network.

\subsection{Limiting History and Future Contexts in Attention}
Attending to the whole input acoustic sequences in attention computations (i.e. full attention) not only disables the streaming inference but also gives the high computational complexity, $O(T^2)$ for computing encoder outputs. To reduce the computational cost and latency, we limit the left history frames ($L$) and future frames ($R$), $(\mathbf{x}_{t-L},...,\mathbf{x}_{t-R})$, of multi-channel encoder to compute $\mathbf{h}_{t_i}$. We also limit the left history frames ($L$) of the label encoder to compute $\mathbf{h}_{u_{i-1}}$. However, it also comes with potential performance drop, as investigated in experiments.

\begin{table}[t]
\begin{center}
\caption{The WERRs (\%) of MCTT over MCT~\cite{chang2021end} for 2-channel inputs, and 3-channel inputs with different combiners.}
\label{tab:over_mct_mcconvt}
\vskip-7pt
\begin{tabular}{l|@{\hskip 1mm}c@{\hskip 1mm}|@{\hskip 1mm}c@{\hskip 1mm}|@{\hskip 1mm}c@{\hskip 1mm}|@{\hskip 1mm}c@{\hskip 1mm}} \hline
 & Model Size &  &  &  \\
Method & (Million) &  combiner & test-clean & test-other  \\ \hline
MCT-2\cite{chang2021end} & 18.59 & - & 0 & 0 \\
MCTT-2 & 17.53 & - & \textbf{11.62} & \textbf{4.51} \\ \hline
MCT-3\cite{chang2021end} & 20.43 & \emph{Affine} & 0 & 0 \\
MCT-3\cite{chang2021end} & 18.59 & \emph{Avg} & 7.01 & 6.99 \\
MCTT-3 & 17.53 & \emph{Avg} & \textbf{11.55} & \textbf{8.32} \\
MCTT-3 & 17.53 & \emph{Concat} & 10.44 & 5.41 \\ 
\hline
\end{tabular}
\end{center}
\vspace{-10pt}
\end{table}

\section{Experiments}
\label{sec:exps}

\subsection{Dataset}
\label{subsec:dataset}
To evaluate our multi-channel transformer transducer (MCTT), we conduct a series of ASR experiments using over 2,200 hours of speech utterances from our in-house de-identified far-field dataset. The amount of training set, validation set (for model hyper-parameter selection), and test set are 2,000 hours, 24 hours, and 233 hours respectively.
The device-directed speech data was captured using a smart speaker with 7 microphones, and a 63 mm aperture. The evaluation set has abundant annotations including the estimated SNR levels, and test-clean (no background speech) as well as test-other (with background speech) splits . In this dataset, 2 microphone signals of aperture distance and the super-directive beamformed signal by \cite{doclo2007superdirective} using 7 microphone signals are employed through all the experiments.

\begin{table}[t]
\begin{center}
\caption{The inference speed comparisons of MCTT and MCT~\cite{chang2021end} in terms of Wall Clock Time (WCT).}
\label{tab:decoding_times}
\vskip-7pt
\begin{tabular}{l|@{\hskip 1mm}c@{\hskip 1mm}|@{\hskip 1mm}c@{\hskip 1mm}} \hline
 & Model Size & WCT (sec) \\
Method & (Million) & TP50\hspace{2.5mm}TP90\hspace{2.5mm}TP99 \\ \hline
MCT\cite{chang2021end} & 18.59 & 4.26\hspace{2.5mm}5.65\hspace{2.5mm}5.91 \\
MCTT & 17.53 & \textbf{0.27}\hspace{2.5mm}\textbf{0.48}\hspace{2.5mm}\textbf{0.74}  \\
\hline
\end{tabular}
\end{center}
\vspace{-15pt}
\end{table}

\subsection{Baselines}
Following~\cite{chang2021end}, one of the baselines is single channel + Transformer Transducer (SC-TT);  we feed each of two raw channels individually into the transformer transducer for training and testing, and pick the best performed one. In addition, we compare to three stagewise beamforming methods cascaded with the transformer transducer (TT) models. The beamforming methods include Super-directive beamformer (SDBF) \cite{doclo2007superdirective}, Neural beamformer (NBF) \cite{kumatani2019multi}, and Neural masked-based beamformer (NMBF) \cite{heymann2016neural}. We denote the stagewise methods as SDBF-TT, NBF-TT, NMBF-TT, respectively. Note that SDBF-TT uses 7 microphone signals for beamforming as mentioned in section~\ref{subsec:dataset} while NBF-TT, NMBF-TT, and the proposed MCTT all take only 2 microphone signals as inputs. 
We also compare our method to multi-channel transformer network (MCT)~\cite{chang2021end}, which is a single integrated multi-channel model.  

\subsection{Experimental Setup and Evaluation Metric}
We set the number of audio encoder layers ($N_{AE}$=12) and label encoder layers ($N_{LE}$=6 for SC-TT, SDBF-TT, NBF-TT, NMBF-TT, $N_{LE}$=4 for MCT and MCTT) with 512 neurons to make all models with comparable number of parameters (18 millions), except for NMBF-TT (25.39 millions) due to the additional mask estimator~\cite{heymann2016neural}. Following \cite{chang2021end}, we use log-STFT square magnitude and phase features~\cite{wang2018combining,wang2018multi} as inputs of our method, which are extracted every 10 ms with a window size of 25 ms from audio samples. The same setting is also applied to the feature extraction for baselines following \cite{chang2021end}. 
The Adam optimizer \cite{kingma2014adam}, and subword tokenizer \cite{sennrich-etal-2016-neural} with $4,001$ tokens are exploited. Results of all the experiments are reported as relative word error rate reduction (WERR)~\cite{chang2021end}. The higher the WERR is the better.

\subsection{Comparisons to Stagewise Multi-channel Models}
We first compare the performance of MCTT with 2 channels, \emph{Avg} combiner (MCTT-2) to the stagewise beamforming plus transformer transducer models, all with full attention audio encoder. The results are illustrated in Fig.~\ref{fig:werrs}. 
As shown in Fig.~\ref{fig:werrs} (a), MCTT-2 outperforms SC-TT by 7.1\% and neural beamformer + acoustic models (NBF-TT and NMBF-TT) by 6\% in average. MCTT-2 also performs better than SDBF-TT by 2.48\% even though it only considers 2 raw channels (2 chs). We further investigate if the super-directive beamformed signal is complementary to the other 2 channels by taking it as the third channel and feed them all to MCTT (denoted as MCTT-3). As can be seen in Fig.~\ref{fig:werrs} (a), it provides 4\% more improvements (WERRs) in average over all baselines as comparing to MCTT-2. In Fig.~\ref{fig:werrs} (b), we further compare different methods w.r.t. different SNR levels. Again, we observe MCTT-2,3 achieve consistent improvements over SC-TT comparing to other methods across different SNRs.

\subsection{Comparisons to Multi-channel Transformer}
Next, we compare the proposed MCTT to MCT~\cite{chang2021end} with 2 channels and 3 channels (2 raw channels plus the super-directive beamformed signal) as inputs with different combiners. They are denoted as MCT-2,3 and MCTT-2,3 respectively. 
Note the combiner introduced in Sec.~\ref{sec:MCEnc} is not needed for the 2-channel case, so its effect is only reported for the 3-channel case.
We observe in Table~\ref{tab:over_mct_mcconvt} that MCTT-2 outperforms MCT-2 especially in test-clean split. Both MCT-3 and MCTT-3 with \emph{Avg} combiner perform better than MCT-3 with \emph{Affine} combiner, and MCTT-3 performs the best. Besides, using \emph{Avg} combiner is more effective than using \emph{Concat} combiner.

We further evaluate inference speed by measuring decoding time over 10,000 utterances on a Intel Xeon® Platinum 8175M processors machine using 1 CPU per method to process an utterance at a time with greedy search decoding. The Top Percentile values, TP50 (median), TP90, and TP99 wall clock times (WCT) are shown in Table~\ref{tab:decoding_times}. 
Most of inference time of MCT has been dedicated to the encoder-decoder attention, while MCTT does not have this issue and achieves $15.4$ times faster inference speed in terms of TP50.

\begin{table}[t]
\begin{center}
\caption{The WERRs (\%) over full-attention MCTT (all contexts=``inf'') by limiting left context per layer for label encoder.}
\label{tab:label_encoder_l}
\vskip-7pt
\begin{tabular}{c@{\hskip 1.2mm}c@{\hskip 1.2mm}|c@{\hskip 1.2mm}|c@{\hskip 1.2mm} | c@{\hskip 1.2mm}} \hline
\multicolumn{2}{c|}{MC Audio Mask} & \multicolumn{1}{c|}{Label Mask} & \multicolumn{2}{c}{WERR (\%)} \\
\hspace{2mm}L & \hspace{2mm}R & L & test-clean & test-other \\ \hline
\hspace{2mm}inf & \hspace{2mm}inf & inf & 0 & 0 \\
\hspace{2mm}inf & \hspace{2mm}inf & 20 & \textbf{2.10} & -0.24 \\
\hspace{2mm}inf & \hspace{2mm}inf & 4 & 1.17 & \textbf{0.95} \\ \hline
\hspace{2mm}inf & \hspace{2mm}10 & inf & -3.27 & \textbf{-3.76} \\
\hspace{2mm}inf & \hspace{2mm}10 & 20 & \textbf{-2.68} & -5.25 \\
\hspace{2mm}inf & \hspace{2mm}10 & 4 & -3.10 & -4.00 \\ \hline
\end{tabular}
\end{center}
\vskip-10pt
\end{table}

\begin{table}[t]
\begin{center}
\caption{The WERRs (\%) over full audio attention based MCTT by limiting right context (R) per layer for MC audio encoder.}
\label{tab:audio_encoder_r}
\vskip-7pt
\begin{tabular}{c@{\hskip 1.2mm}c@{\hskip 1.2mm}|c@{\hskip 1.2mm}|c@{\hskip 1.2mm} | c@{\hskip 1.2mm}} \hline
\multicolumn{2}{c|}{MC Audio Mask} & \multicolumn{1}{c|}{Label Mask} & \multicolumn{2}{c}{WERR (\%)} \\
\hspace{2mm}L & \hspace{2mm}R & L & test-clean & test-other \\ \hline
\hspace{2mm}inf & \hspace{2mm}inf & 20 & 0 & 0 \\
\hspace{2mm}inf & \hspace{2mm}0 & 20 & -23.65 & -16.56 \\
\hspace{2mm}inf & \hspace{2mm}2 & 20 & -12.17 & -8.04 \\
\hspace{2mm}inf & \hspace{2mm}6 & 20 & -5.99 & \textbf{-2.74} \\
\hspace{2mm}inf & \hspace{2mm}10 & 20 & \textbf{-4.88} & \textbf{-5.00} \\
\hline
\end{tabular}
\end{center}
\vskip-10pt
\end{table}

\begin{table}[t]
\begin{center}
\caption{The WERRs (\%) over full audio attention based MCTT (with left context of label encoder L=20) by limiting both MC audio contexts (L and R) and label contexts (L) for streaming.}
\label{tab:context_l_r}
\vskip-7pt
\begin{tabular}{c@{\hskip 1.2mm}c@{\hskip 1.2mm}|c@{\hskip 1.2mm}|c@{\hskip 1.2mm} | c@{\hskip 1.2mm}} \hline
\multicolumn{2}{c|}{MC Audio Mask} & \multicolumn{1}{c|}{Label Mask} & \multicolumn{2}{c}{WERR (\%)} \\
\hspace{2mm}L & \hspace{2mm}R & L & test-clean & test-other \\ \hline
\hspace{2mm}inf & \hspace{2mm}inf & 20 & 0 & 0 \\
\hspace{2mm}20 & \hspace{2mm}0 & 20 & -27.68 & -22.87 \\
\hspace{2mm}20 & \hspace{2mm}10 & 20 & \textbf{-7.37} & \textbf{-6.79} \\
\hspace{2mm}20 & \hspace{2mm}20 & 20 & \textbf{-7.54} & \textbf{-6.31} \\ \hline
\hspace{2mm}10 & \hspace{2mm}0 & 20 & -24.08 & -22.45 \\
\hspace{2mm}10 & \hspace{2mm}10 & 20 & -9.51 & -11.08 \\
\hspace{2mm}10 & \hspace{2mm}20 & 20 & \textbf{-7.28} & \textbf{-7.92} \\ \hline
\end{tabular}
\end{center}
\vskip-10pt
\end{table}

\subsection{Results of Limiting Contexts in Attention Computation}
Finally, we ran training and decoding experiments using MCTT with limited attention windows over audio and text labels, with a view to build streaming multi-channel (MC) speech recognition systems with low latency and low computation cost. ``$\text{inf}$" in Table~\ref{tab:label_encoder_l},~\ref{tab:audio_encoder_r},~\ref{tab:context_l_r} means we employ all of the left or right contexts. Besides, MC Audio Mask, and Label Mask indicate the coverage of audio/label frames to be considered in attention of Multi-Channel audio encoder and label encoder respectively.

We start from evaluating how the left context of the label encoder affects performance. In Table~\ref{tab:label_encoder_l}, we show that constraining each layer to use only 4 previous label frames yields the similar accuracy with the model using all previous frames per layer ($1.06\%$ WERR in average when MC audio mask R=inf). As constraining right context of MC audio to 10, the WERR differences are also small; the maximum WERR difference is $0.24\%$ (-$3.76$\%-(-$4$)\%) when compared to using all previous frames per layer. It indicates that very limited left context for label encoder is good enough for MCTT.

We then fix the left context of label encoder to 20
, and constrain the MC audio encoder to attend to only the left of the current frame (so that no latency is introduced). As shown in Table~\ref{tab:audio_encoder_r}, the WERs drastically degrade by $23.65\%$ and $16.56\%$ in test-clean and test-other splits comparing to MCTT with full attention MC audio encoder. By allowing the model to see some future frames (e.g. $R=10$), we can bring down the WER degradation to $\leq 5\%$ for both splits.
 
Table~\ref{tab:context_l_r} reports the results when limiting both the left and right contexts of MC audio encoder. By doing so, not only the latency can be reduced, but also the time complexity for one-step inference becomes a constant. We limit the left context of MC audio encoder to 20 and 10 respectively, and then increase right context from 0 to 20. As can be seen in both cases, with the look-ahead to few future frames (e.g. $R=20$), the WER gap to the full-attention audio encoder based model was narrowed down to $6.31\%$ and $7.92\%$ respectively in test-other split.

\section{Conclusion}
\label{sec:conc}
We propose a novel speech recognition model, Multi-Channel Transformer Transducer, which is capable of leveraging multi-channel inputs in an end-to-end fashion and applicable to streaming decoding for speech recognition. We show that the proposed MCTT outperforms its stagewise counterparts, and significantly reduces the inference time against multi-channel transformer~\cite{chang2021end}. Furthermore, by limiting the left contexts and with look-ahead to few future frames, we can not only improve the computation cost, but also bridge the gap between the performance of left-only attention and full attention models.

\bibliographystyle{IEEEtran}
\bibliography{references}

\end{document}